\documentclass[aps,preprint]{revtex4}%
\usepackage{amsfonts}
\usepackage{amsmath}
\usepackage{amssymb}
\usepackage{graphicx}%
\begin{document}
\title{Black Hole solutions in Einstein-Maxwell-Yang-Mills-Gauss-Bonnet Theory }
\author{S. Habib Mazharimousavi$^{\ast}$}
\author{M. Halilsoy$^{\dag}$}
\affiliation{Department of Physics, Eastern Mediterranean University,}
\affiliation{G. Magusa, north Cyprus, Mersin-10, Turkey}
\affiliation{$^{\ast}$habib.mazhari@emu.edu.tr}
\affiliation{$^{\dagger}$mustafa.halilsoy@emu.edu.tr}

\begin{abstract}
We consider Maxwell and Yang-Mills (YM) fields together, interacting through
gravity both in Einstein and Gauss-Bonnet (GB) theories. For this purpose we
choose two different sets of Maxwell and metric ansaetze. In our first ansatz,
asymptotically for $r\rightarrow0$ (and $N>4$) the Maxwell field dominates
over the YM field. In the other asymptotic region, $r\rightarrow\infty$,
however, the YM field becomes dominates. For $N=3$ and $N=4$, where the GB
term is absent, we recover the well-known Ba\~{n}ados-Teitelboim-Zanelli (BTZ)
and Reissner-Nordstr\"{o}m (RN) metrics, respectively. The second ansatz
corresponds to the case of constant radius function for $S^{N-2}$ part in the
metric. This leads to the maximally symmetrical, everywhere regular
Bertotti-Robinson (BR) type solutions to represent our cosmos both at large
and small.

\end{abstract}
\maketitle

\section{INTRODUCTION}

We introduce Maxwell field alongside with Yang-Mills (YM) field in general
relativity and present spherically symmetric black hole solutions in any
higher dimensions. These two gauge fields, one Abelian the other non-Abelian,
are coupled through gravity and to our knowledge they have not been considered
together in a common geometry so far. Being linear, a multitude of Maxwellian
fields can be superimposed at equal ease, but for the YM field there is no
such freedom. Our treatment of YM field in this paper like Maxwell field is
completely classical (i.e. non-quantum). From physics standpoint,
electromagnetism has long range effects and dominates outside the nuclei of
natural matter. YM field on the other hand is confined to act inside nuclei,
however, the existence of exotic and highly dense matter in our universe
encourages us to use YM field in a broader sense. To obtain exact solutions
the Maxwell field is chosen pure electric while the YM field is pure magnetic.
Historically, starting with the Reissner-Nordstr\"{o}m (RN) metric, its higher
dimensional extensions known as the Einstein-Maxwell (EM) black holes are
well-known by now. Einstein-Yang-Mills (EYM) black holes in higher dimensions
have also attracted interest in more recent works \cite{1,2,3}. In this paper
we combine these two foregoing black holes (i.e. EM and EYM) under the common
title of EMYM black holes. Further, we add to it the Gauss-Bonnet (GB) term
overall which we phrase as the EMYMGB black holes\cite{17}(and the references
therein). Such black holes will be characterized by mass ($M$), Maxwell charge
($q$), YM charge ($Q$), GB parameter $\alpha$ and cosmological constant
$\left(  \Lambda\right)  .$ For a broad class of black hole solutions we
investigate the thermodynamics and other properties as manifestations of these
physical parameters. All these parameters naturally have imprints an planetary
motion, gravitational lensing, ripping apart of stars and ultimately on the
accurate picture of our cosmos. It is remarkable that exact solutions to such
a highly non-linear theory can be found for two different Maxwell and metric
ansaetze. Before adding the GB term we find the EMYM solutions in which the
relative contributions from the Maxwell /YM charges can be compared. It is
shown that for $r\rightarrow0$ Maxwell term dominates over the YM term which
is reminiscent of asymptotic freedom from the YM charge. There is no need to
remind that the latter is a quantum effect while our treatment is entirely
classical here. For $r\rightarrow\infty$ the YM term becomes dominant.
Although this may sound contradictory to the short rangeness of the YM field
we may attribute this to the decisive role played by the higher dimensionality
of spacetime.\ Such behaviors, we speculate, have relevance in connection with
the mini (super-massive) black holes as well as asymptotically (anti)-de
Sitter space times. The solutions for $N>5$ turn out to be distinct from the
lower (i.e., $N=3,4$) dimensional cases. We show that in the latter cases of
$N=3$ and $N=4$ the role of a YM charge is similar to the Maxwell charge as
far as the space time metric is concerned. Namely, one is still the BTZ, while
the other is the RN metric. It is found that for $N=3$ and $N=5$ the metric
function contains a logaritmic term while in the other cases we have inverse
power low dependence on $r$. The unprecedented logarithmic potential lead to
entirely different Keplerian orbits. Since black holes are scale invariant
objects, i. e., they occur both at micro/ macro scales as mini/ supermassive
black holes, the orbits revealed in one scale is valid in all scales. To show
this we investigate the Newtonian approximation (for $N=5$) as projected into
the polar plane which reveals the role of the YM charge in forming bound
states with smaller orbits. Our second ansatz leads, beside the black hole
solution, to a Bertotti-Robinson (BR) type metric within the EMYMGB theory. It
is known that extremal RN, which is supersymmetric soliton solution to
extended supergravity, interpolates between two maximally symmetric spacetime,
namely the flat space at infinity and BR at the horizon $\left(  r=0\right)
$. In other words the near horizon geometry of an extremal RN black hole is
identified as the BR spacetime. This idea can be extended to higher dimensions
as p-branes. The topology of N-dimensional BR type YM field is, as in the EM
theory, $adS_{2}\times S^{N-2}$ (i.e., anti de Sitter $\times$ ($N-2$) sphere)
with a marked difference in their radii. The maximal symmetry of the BR
spacetime, without singularity makes it a favorable model to represent our
homogenous and isotropic universe in the absence of rotation. For the
representative of the rotating BR universe and its cosmological implications
in $N=4$ we refer elsewhere [19]. The interesting feature here is that the BR
parameter consist of all the parameters of the theory, namely, $M,$ $q,$ $Q,$
$\Lambda$ and $\alpha.$ In other words, even the topological GB parameter
serves to construct a finely- tuned cosmological BR model with maximal
symmetry. \bigskip It was shown by Gibbons and Townsend \cite{1} that extreme
self-gravitating Yang-type monopole also yields an $adS_{2}\times S^{2N}$
topological vacuum in the $\left(  2N+2\right)  $ dimensional EYM theory. It
is our belief that our metrics will be useful both in string/ supergravity
theories as well as in cosmology. One immediate conclusion we can draw is that
we can generate an "effective cosmological constant $\Lambda_{eff}$ " to play
the same role, in the absence of a real $\Lambda.$ To feel confident, we
verify also that our physical sources in higher dimensions satisfy all the
weak, strong, dominant and causality conditions. 

The paper is organized as follows: In Sec. ($II$) we introduce our action,
metric, Maxwell, YM ansatz and the field equations. Sec. ($III$) follows with
the solution of the field equations for different dimensionalities. The
geometry outside a 5-dimensional black hole with its Newtonian approximation
is investigated in Sec. ($IV$). In Sec. ($V$) we add the Gauss-Bonnet (GB)
term and present the most general solution. Sec. ($VI$) deals with a new set
of ansatz for the Maxwell field and metric function in which we study the
black hole properties and BR classes inherent in them. Energy and causality
conditions are discussed in Sec. ($VII$). The paper ends with concluding
remarks in Sec. ($VIII$).

\section{Action, Field Equations and our ansaetze (RN type)}

The action which describes Einstein-Maxwell-Yang-Mills gravity with a
cosmological constant in $N$ dimensions reads
\begin{equation}
I_{G}=\frac{1}{2}\int_{\mathcal{M}}dx^{N}\sqrt{-g}\left(  R-\frac{\left(
N-1\right)  \left(  N-2\right)  }{3}\Lambda-F_{\mu\nu}F^{\mu\nu}%
-\mathbf{Tr}\left(  F_{\mu\nu}^{(a)}F^{(a)\mu\nu}\right)  \right)
\end{equation}
where in this context we use the following abbreviation,
\begin{equation}
\mathbf{Tr}\left(  .\right)  =\overset{\left(  N-1\right)  (N-2)/2}%
{\underset{a=1}{\sum}}\left(  .\right)  .
\end{equation}
Here, $R$ is the Ricci Scalar while the YM and Maxwell fields are defined
respectively by
\begin{align}
F_{\mu\nu}^{\left(  a\right)  } &  =\partial_{\mu}A_{\nu}^{\left(  a\right)
}-\partial_{\nu}A_{\mu}^{\left(  a\right)  }+\frac{1}{2\sigma}C_{\left(
b\right)  \left(  c\right)  }^{\left(  a\right)  }A_{\mu}^{\left(  b\right)
}A_{\nu}^{\left(  c\right)  },\\
F_{\mu\nu} &  =\partial_{\mu}A_{\nu}-\partial_{\nu}A_{\mu},\nonumber
\end{align}
in which $C_{\left(  b\right)  \left(  c\right)  }^{\left(  a\right)  }$
stands for the structure constants of $\frac{\left(  N-1\right)  (N-2)}{2}$
parameter Lie group $G$ and $\sigma$ is a coupling constant. $A_{\mu}^{\left(
a\right)  }$ are the $SO(N-1)$gauge group YM potentials while $A_{\mu}$
represents the usual Maxwell potential. We note that the internal indices
$\{a,b,c,...\}$ do not differ whether in covariant or contravariant form.
Variation of the action with respect to the space-time metric $g_{\mu\nu}$
yields the field equations
\begin{equation}
G_{\mu\nu}+\frac{\left(  N-1\right)  \left(  N-2\right)  }{6}\Lambda g_{\mu
\nu}=T_{\mu\nu},
\end{equation}
where the stress-energy tensor is the superposition of the Maxwell and YM
parts, namely
\begin{equation}
T_{\mu\nu}=\left(  2F_{\mu}^{\lambda}F_{\nu\lambda}-\frac{1}{2}F_{\lambda
\sigma}F^{\lambda\sigma}g_{\mu\nu}\right)  +\mathbf{Tr}\left[  2F_{\mu
}^{\left(  a\right)  \lambda}F_{\nu\lambda}^{\left(  a\right)  }-\frac{1}%
{2}F_{\lambda\sigma}^{\left(  a\right)  }F^{\left(  a\right)  \lambda\sigma
}g_{\mu\nu}\right]  .
\end{equation}
Variation with respect to the gauge potentials $A_{\mu}^{\left(  a\right)  }$
and $A_{\mu}$ yield the respective YM and Maxwell equations%
\begin{equation}
F_{;\mu}^{\left(  a\right)  \mu\nu}+\frac{1}{\sigma}C_{\left(  b\right)
\left(  c\right)  }^{\left(  a\right)  }A_{\mu}^{\left(  b\right)  }F^{\left(
c\right)  \mu\nu}=0,\text{ \ \ }F_{;\mu}^{\mu\nu}=0,
\end{equation}
whose integrability equations in order are%
\begin{equation}
^{\ast}F_{;\mu}^{\left(  a\right)  \mu\nu}+\frac{1}{\sigma}C_{\left(
b\right)  \left(  c\right)  }^{\left(  a\right)  }A_{\mu}^{\left(  b\right)
\ast}F^{\left(  c\right)  \mu\nu}=0,\text{ \ \ }^{\ast}F_{;\mu}^{\mu\nu}=0,
\end{equation}
in which $^{\ast}$ means duality \cite{4}. The N-dimensional spherically
symmetric line element is chosen as%
\begin{equation}
ds^{2}=-f(r)\;dt^{2}+\frac{dr^{2}}{f(r)}+r^{2}d\;\Omega_{N-2}^{2},
\end{equation}
in which the $S^{N-2}$ line element will be expressed in the standard
spherical form%
\begin{equation}
d\Omega_{N-2}^{2}=d\theta_{1}^{2}+\underset{i=2}{\overset{N-2}{%
{\textstyle\sum}
}}\underset{j=1}{\overset{i-1}{%
{\textstyle\prod}
}}\sin^{2}\theta_{j}\;d\theta_{i}^{2},
\end{equation}
where%
\[
0\leq\theta_{N-2}\leq2\pi,0\leq\theta_{i}\leq\pi,\text{ \ \ }1\leq i\leq N-3.
\]
For the YM field we employ the magnetic Wu-Yang ansatz \cite{3,5,6} where the
potential 1-forms are expressed by%
\begin{align}
\mathbf{A}^{(a)} &  =\frac{Q}{r^{2}}\left(  x_{i}dx_{j}-x_{j}dx_{i}\right)
,\text{ \ \ }Q=\text{charge, \ }r^{2}=\overset{N-1}{\underset{i=1}{\sum}}%
x_{i}^{2},\\
2 &  \leq j+1\leq i\leq N-1,\text{ \ and \ }1\leq a\leq\left(  N-1\right)
(N-2)/2,\nonumber
\end{align}
The Maxwell potential 1-form is chosen as
\begin{equation}
\mathbf{A}=%
\genfrac{\{}{.}{0pt}{0}{\frac{q}{r^{N-3}}dt,\text{ \ \ \ }N\geq4}{q\ln\left(
r\right)  dt,\text{ \ \ }N=3}%
\end{equation}
for the electric charge $q$ .The following sections will be devoted to the
EMYM equations (i.e. Eq. (4)) and their solutions for all dimensionalities
$N\geq3$. In the last two sections we add GB theory (for $N\geq5$) into our
formalism and find new combined solutions for different Maxwell and metric
ansaetze. The energy-momentum tensor for the Maxwell and YM fields for
$N\geq4$ (3-dimensional case will be studied separately) are given by \cite{3}%
\begin{equation}
T_{\text{Max }b}^{a}=-\frac{\left(  N-3\right)  ^{2}q^{2}}{r^{2\left(
N-2\right)  }}\text{diag}\left[  1,1,-1,-1,..,-1\right]  ,
\end{equation}%
\begin{align}
T_{\text{YM }b}^{a} &  =-\frac{\left(  N-3\right)  \left(  N-2\right)  Q^{2}%
}{2r^{4}}\text{diag}\left[  1,1,\kappa,\kappa,..,\kappa\right]  ,\\
\kappa &  =\frac{N-6}{N-2}.\nonumber
\end{align}

\section{EMYM solution for $N\geq6$}

The basic field equation that incorporates all relevant expressions is given by%

\begin{equation}
\frac{r^{3}}{N-3}\left(  f^{\prime}+\frac{\Lambda}{3}\left(  N-1\right)
r\right)  +\left(  f-1\right)  r^{2}+Q^{2}+\frac{2q^{2}\left(  N-3\right)
r^{2\left(  4-N\right)  }}{\left(  N-2\right)  }=0,
\end{equation}
where $f^{\prime}\equiv\frac{df}{dr}$ ,which admits the integral%

\begin{equation}
f\left(  r\right)  =1-\frac{2M}{r^{N-3}}-\frac{\Lambda}{3}r^{2}+\frac{\left(
N-3\right)  2q^{2}}{\left(  N-2\right)  r^{2\left(  N-3\right)  }}%
-\frac{\left(  N-3\right)  Q^{2}}{\left(  N-5\right)  r^{2}},
\end{equation}
for the constant of integration $M$ as the mass parameter.

It is observed that this solution is not valid for $N=5$ and $N=3,$ for these
particular cases therefore different solutions will be found. It is valid,
however, for $N\geq6,$ which implies that the signs of Maxwell and YM terms
are opposite. This may have interesting consequences pertaining to the
confinement of a system that possesses both type of charges. We express (for
$\Lambda=0$)%
\begin{equation}
f\left(  r\right)  =1+2\Phi\left(  r\right)
\end{equation}
where $\Phi\left(  r\right)  $ stands for the Newtonian-like potential, which
we identify as%

\begin{equation}
\Phi\left(  r\right)  =-\frac{M}{r^{N-3}}+\frac{\left(  N-3\right)  q^{2}%
}{\left(  N-2\right)  r^{2\left(  N-3\right)  }}-\frac{\left(  N-3\right)
Q^{2}}{2\left(  N-5\right)  r^{2}}.
\end{equation}
We can define the active force through $F=-\mathbf{\nabla}\Phi,$ which yields%

\begin{equation}
F\left(  r\right)  =-\frac{M\left(  N-3\right)  }{r^{\left(  N-2\right)  }%
}+\frac{2q^{2}\left(  N-3\right)  ^{2}}{\left(  N-2\right)  r^{2N-5}}%
-\frac{\left(  N-3\right)  Q^{2}}{\left(  N-5\right)  r^{3}}.
\end{equation}
The signs of the Maxwell and YM terms reveal that while the former is
repulsive the latter becomes attractive. One can easily show that for
$r\rightarrow\infty$ the YM term dominates (let $\Lambda=0$), namely
\begin{equation}
\underset{r\rightarrow\infty}{\lim}f\left(  r\right)  \rightarrow
1-\frac{\left(  N-3\right)  Q^{2}}{\left(  N-5\right)  r^{2}}.
\end{equation}
For $r\rightarrow0^{+}$ we have the opposite case,%
\begin{equation}
\underset{r\rightarrow0^{+}}{\lim}f\left(  r\right)  \rightarrow
1+\frac{\left(  N-3\right)  2q^{2}}{\left(  N-2\right)  r^{2\left(
N-3\right)  }},
\end{equation}
which may be interpreted as an "asymptotic independence " from one type of
charge (or the other )in different limits. For the mini black holes this has
the striking effect that the Hawking temperature depends only on the electric charge.

To find the radius of possible horizon(s), we equate the metric function
$f\left(  r\right)  $ to zero, which leads to the following equation%
\begin{gather}
6\left(  N-3\right)  \left(  N-5\right)  q^{2}-6M\left(  N-2\right)  \left(
N-5\right)  r^{N-3}-\Lambda\left(  N-2\right)  \left(  N-5\right)  r^{2\left(
N-2\right)  }+\\
3\left(  N-2\right)  \left(  N-5\right)  r^{2\left(  n-3\right)  }-3\left(
N-2\right)  \left(  N-3\right)  r^{2\left(  N-4\right)  }=0.\nonumber
\end{gather}
This equation in six dimensional spacetime, without the cosmological constant
and zero mass has an exact real solution%
\begin{equation}
r_{h}=\sqrt{\frac{1}{6}\left(  \frac{1}{2}\sqrt[3]{\Delta}+\frac{2\tilde
{Q}^{4}}{\sqrt[3]{\Delta}}+\tilde{Q}^{2}\right)  }%
\end{equation}
where%
\begin{align}
\Delta &  =8\tilde{Q}^{6}-108\tilde{q}^{2}+12\sqrt{3\tilde{q}^{2}\left(
27\tilde{q}^{2}-4\tilde{Q}^{6}\right)  },\\
\tilde{Q}^{2}  &  =6Q^{2},\text{ \ \ }\tilde{q}^{2}=12q^{2},\text{ \ \ }%
\tilde{q}^{2}\geq\frac{4\tilde{Q}^{6}}{27}.\nonumber
\end{align}
\qquad

\subsection{The case N=5}

For $N=5$, the master equation (14) admits the solution%

\begin{equation}
f\left(  r\right)  =1-\frac{2M}{r^{2}}-\frac{\Lambda}{3}r^{2}+\frac{4q^{2}%
}{3r^{4}}-\frac{2Q^{2}\ln\left(  r\right)  }{r^{2}},
\end{equation}
which involves an unusual logarithmic term. This is an asymptotically flat
black hole solution in which finding the exact radius of the horizon of the
possible black holes, leads us to the following non-algebric equation%
\begin{equation}
\frac{\Lambda}{3}r^{6}-r^{4}+2Mr^{2}+2Q^{2}r^{2}\ln r-\frac{4}{3}q^{2}=0.
\end{equation}
This equation, even without cosmological constant does not give an analytical
solution. By using numerical method, we plot the root ($r_{h})$ of the above
equation in Fig. (1), to show the contribution of the Maxwell and YM charges
to the construction of such possible black holes. The Hawking temperature
$T_{H}$ can be written as%
\begin{equation}
T_{H}=\frac{\kappa}{2\pi}=\frac{1}{4\pi}\left\vert f^{\prime}\left(
r_{h}\right)  \right\vert =\frac{1}{4\pi}\left\vert \frac{1}{r_{h}}-\frac
{2}{3}\Lambda r_{h}-\frac{4}{3}\frac{q^{2}}{r_{h}^{5}}-\frac{Q^{2}}{r_{h}^{3}%
}\right\vert
\end{equation}
where $r_{h}$ is the radius of the event horizon and $\kappa$ stands for the
surface gravity. The corresponding Newtonian-like YM force term in this case
has the form%
\begin{equation}
force\sim\frac{Q^{2}}{r^{3}}\left(  1-2\ln r\right)  ,
\end{equation}
which implies that it is attractive (repulsive) for $r>\sqrt{e}$ ($r<\sqrt{e}%
$).The Maxwell term remains always repulsive. We notice also that for
$r\rightarrow\infty$ (for $\Lambda=0$) the YM term dominates over the Maxwell%

\begin{equation}
\underset{r\rightarrow\infty}{\lim}f\left(  r\right)  \rightarrow
1-\frac{2Q^{2}\ln\left(  r\right)  }{r^{2}}.
\end{equation}
In the limit $r\rightarrow0^{+}$, on the other hand we obtain
\begin{equation}
\underset{r\rightarrow0^{+}}{\lim}f\left(  r\right)  \rightarrow1+\frac
{4q^{2}}{3r^{4}},
\end{equation}
which is in confirm with the behavior (20) for $N\geq6.$

\subsection{The case N=4}

In 4-dimensional case the solution for the metric function $f\left(  r\right)
$ is given by
\begin{equation}
f\left(  r\right)  =1-\frac{2M}{r}-\frac{\Lambda}{3}r^{2}+\frac{\left(
q^{2}+Q^{2}\right)  }{r^{2}},
\end{equation}
in which both the Maxwell and YM charges have similar feature and the metric
is the well-known RN de-Sitter. Extremal YM black hole for instance, follows
in analogy with the 4-dimensional RN black hole. Both charges act repulsively
against the attractive property of mass. The black hole solution for $f\left(
r\right)  =0$ given in Eq. (30) without cosmological constant, has two roots
\begin{equation}
r_{\pm}=M\pm\sqrt{M^{2}-\left(  q^{2}+Q^{2}\right)  }%
\end{equation}
in which the mass and charges must satisfy the constraint $M^{2}\geq\left(
q^{2}+Q^{2}\right)  $ to have the event $\left(  r_{+}\right)  $ and Cauchy
$\left(  r_{-}\right)  $ horizons. Here the thermodynamic properties of the
solution (30) is exactly same as the four dimensional RN black holes and
therefore we just comment that the roles of the YM and Maxwell charges in the
metric are not distinguished from each other.

\subsection{The case N=3}

In 3-dimensional space time we adopt the Maxwell potential 1-form \cite{7}
\begin{equation}
A=q\ln\left(  r\right)  dt,\text{ \ \ (}q=\text{electric charge),}%
\end{equation}
and introduce the YM gauge potential 1-forms accordingly as
\begin{align}
A^{\left(  1\right)  } &  =Q\cos\left(  \phi\right)  \ln\left(  r\right)
dt,\\
A^{\left(  2\right)  } &  =Q\sin\left(  \phi\right)  \ln\left(  r\right)
dt,\nonumber\\
A^{\left(  3\right)  } &  =-Qd\phi,\text{ \ \ }(Q=\text{YM charge),}\nonumber
\end{align}
which satisfy the Maxwell and YM equations, respectively. The corresponding
EMYM equation for $f\left(  r\right)  ,$ independent from Eq. (14) becomes%

\begin{equation}
3rf^{\prime}+2\Lambda r^{2}+6\left(  Q^{2}+q^{2}\right)  =0,
\end{equation}
which is readily integrated as
\begin{equation}
f\left(  r\right)  =-M-\frac{1}{3}\Lambda r^{2}-2\left(  q^{2}+Q^{2}\right)
\ln\left(  r\right)  ,
\end{equation}
and the line element is
\begin{equation}
ds^{2}=-f\left(  r\right)  dt^{2}+\frac{dr^{2}}{f\left(  r\right)  }%
+r^{2}d\theta^{2}.
\end{equation}
A negative cosmological constant leads to a black hole solution \cite{7}%
\begin{equation}
f\left(  r\right)  =-M+\frac{1}{3}\left\vert \Lambda\right\vert r^{2}-2\left(
q^{2}+Q^{2}\right)  \ln\left(  r\right)
\end{equation}
in which the possible radii of the horizons are given by%
\begin{align}
r_{+}  &  =\exp\left[  -\frac{1}{2}LamberW\left(  -1,\frac{-\left\vert
\Lambda\right\vert e^{\frac{-M}{Q^{2}+q^{2}}}}{3\left(  Q^{2}+q^{2}\right)
}\right)  -\frac{M}{2\left(  Q^{2}+q^{2}\right)  }\right] \\
r_{-}  &  =\exp\left[  -\frac{1}{2}LamberW\left(  0,\frac{-\left\vert
\Lambda\right\vert e^{\frac{-M}{Q^{2}+q^{2}}}}{3\left(  Q^{2}+q^{2}\right)
}\right)  -\frac{M}{2\left(  Q^{2}+q^{2}\right)  }\right]
\end{align}
in which $LambertW\left(  k,x\right)  $ stands for the Lambert function
\cite{8}. The energy density, given by%
\begin{equation}
\epsilon=T^{tt}=\frac{Q^{2}+q^{2}}{r^{2}}%
\end{equation}
may be used to calculate the total energy of the black hole. This shows that
the energy diverges logarithmically. The surface gravity, $\kappa$ defined by
\begin{equation}
\kappa^{2}=\left(  -\frac{1}{4}g^{tt}g^{ij}g_{tt,i}g_{tt,j}\right)  _{r=r_{h}%
}=\left(  \frac{1}{2}f^{\prime}\left(  r\right)  \right)  _{r=r_{+}}^{2}%
\end{equation}
gives%
\begin{equation}
\kappa=\left\vert \frac{\left\vert \Lambda\right\vert }{3}r_{+}-\frac
{Q^{2}+q^{2}}{r_{+}}\right\vert .
\end{equation}
Finally we find the Hawking temperature at event horizon as
\begin{equation}
T_{H}=\frac{\kappa}{2\pi}=\frac{1}{2\pi}\left\vert \frac{\left\vert
\Lambda\right\vert }{3}r_{+}-\frac{Q^{2}+q^{2}}{r_{+}}\right\vert
\end{equation}
and a plot of $T_{H}$ is given in Fig. (2) in terms of the mass, $\Lambda$ and
$\left(  Q^{2}+q^{2}\right)  .$

In analogy with the $N=4$ case the squared charges are simply superposed, and
the metric is still the BTZ metric. It is observed that for $q=0\neq Q$ the
same (BTZ) metric describes an EYM black hole in 3-dimensions. Addition of
rotation to the metric, which is beyond our scope here, may add new features
to differ for the Maxwell and YM fields.

\section{The Geometry Outside the 5-dimensional EMYM black hole}

It is evident from solution (24) that, $\xi^{\alpha}=\left(  1,0,0,0,0\right)
$ and $\eta^{\alpha}=$ $\left(  0,0,0,0,1\right)  $ are two of the Killing
vectors associated with the symmetry under displacements in the direction t,
and rotation angle $\psi.$ Accordingly the conserved quantities may be written
as%
\begin{align}
e  &  =-\mathbf{\xi\cdot u=-}g_{\alpha\beta}\xi^{\alpha}u^{\beta}=f\left(
r\right)  u^{t}\\
\ell &  =\mathbf{\eta\cdot u=}g_{\alpha\beta}\eta^{\alpha}u^{\beta}=r^{2}%
\sin^{2}\left(  \theta\right)  \sin^{2}\left(  \phi\right)  u^{\psi}%
\end{align}
where $\mathbf{u=}\left(  u^{t},u^{r},u^{\theta},u^{\phi},u^{\psi}\right)  $
is the five-velocity while $e$ and $\ell$ are the energy density and angular
momentum per unit mass, respectively. We restrict the particle to stay on the
plane $\theta=\frac{\pi}{2},$ $\phi=\frac{\pi}{2}$ with $u^{\theta}=u^{\phi
}=0,$ such that by applying $\mathbf{u\cdot u=}-1$ for the timelike geodesics
one obtains
\begin{equation}
g_{\alpha\beta}u^{\alpha}u^{\beta}=-f\left(  r\right)  \left(  u^{t}\right)
^{2}+\frac{1}{f\left(  r\right)  }\left(  u^{r}\right)  ^{2}+r^{2}\left(
u^{\psi}\right)  ^{2}=-1
\end{equation}
in which $u^{t}=\frac{dt}{d\tau},$ $u^{r}=\frac{dr}{d\tau},$ $u^{\theta}%
=\frac{d\theta}{d\tau},$ $u^{\phi}=\frac{d\phi}{d\tau},$ $u^{\psi}=\frac
{d\psi}{d\tau}$ and $\tau$ is the proper time measured by the observer moving
with the particle. Putting (44) and (45) into (46), one gets%
\begin{equation}
\frac{1}{2}\left(  \frac{dr}{d\tau}\right)  ^{2}+\frac{1}{2}\left[  \left(
\frac{\ell^{2}}{r^{2}}+1\right)  f\left(  r\right)  -1\right]  =\frac{e^{2}%
-1}{2}%
\end{equation}
or equivalently
\begin{equation}
\frac{1}{2}\left(  \frac{dr}{d\tau}\right)  ^{2}+V_{eff}\left(  r\right)
=\mathcal{E}%
\end{equation}
where $\mathcal{E}$ is the density of total energy per unit mass,
$V_{eff}\left(  r\right)  $ is the effective potential for radial motion of
the particle and $f\left(  r\right)  $ is given in (24). In Fig. (3) we plot
$V_{eff}\left(  r\right)  $ in terms of $r$ for different values of $q,$ $Q$
and $\ell$ but zero value for $M.$ Since the particle is restricted to remain
in a plane which only $r$ and $\psi$ would change, we rewrite the equation
(48) in terms of $r$ and $\psi$ as follows:%
\begin{equation}
\frac{1}{2}\left(  \frac{d\psi}{d\tau}\right)  ^{2}\left(  \frac{dr}{d\psi
}\right)  ^{2}+V_{eff}\left(  r\right)  =\mathcal{E}%
\end{equation}
or equivalently from Eq. (49)
\begin{align}
\frac{dr}{d\psi}  &  =\pm\sqrt{\frac{2r^{4}}{\ell^{2}}\left(  \mathcal{E-}%
V_{eff}\left(  r\right)  \right)  },\\
V_{eff}\left(  r\right)   &  =\frac{1}{2}\left[  \left(  \frac{\ell^{2}}%
{r^{2}}+1\right)  f\left(  r\right)  -1\right]  ,\nonumber
\end{align}
in which $\pm$ depends to the initial direction of the motion and we set it to
be positive. In the following subsection, instead of studying the exact
geodesics equation (50), we shall restrict ourselves to the relatively simpler
Newtonian approximation.

\subsection{Newtonian motion}

In the weak field limit, (and setting $\Lambda$ to be zero) one may use
$f\left(  r\right)  =1+2\Phi\left(  r\right)  $ which implies
\begin{equation}
\Phi\left(  r\right)  =-\frac{M}{r^{2}}+\frac{2q^{2}}{3r^{4}}-\frac{Q^{2}%
\ln\left(  r\right)  }{r^{2}},
\end{equation}
and consequently the radial force on a unit mass particle is given by%
\begin{equation}
F_{r}=-\frac{d}{dr}\Phi\left(  r\right)  =-\left(  \frac{2M}{r^{3}}%
-\frac{8q^{2}}{3r^{5}}-\frac{Q^{2}}{r^{3}}+\frac{2Q^{2}\ln\left(  r\right)
}{r^{3}}\right)  .
\end{equation}
The radial equation of motion, therefore, may be written as%
\begin{equation}
\frac{d^{2}r}{dt^{2}}=\frac{\ell^{2}}{r^{3}}-\left(  \frac{2M}{r^{3}}%
-\frac{8q^{2}}{3r^{5}}-\frac{Q^{2}}{r^{3}}+\frac{2Q^{2}\ln\left(  r\right)
}{r^{3}}\right)  ,
\end{equation}
where $\ell$ is the angular momentum per unit mass. As usual, one may start
with the following substitution
\begin{equation}
u=\frac{1}{r}%
\end{equation}
to reduce the last equation into the form
\begin{equation}
\frac{d^{2}u}{d\psi^{2}}+\left(  1+\frac{Q^{2}-2M}{\ell^{2}}\right)
u+\frac{2Q^{2}}{\ell^{2}}u\ln u-\frac{8q^{2}}{3\ell^{2}}u^{3}=0.
\end{equation}
As a particular case we set $1+\frac{Q^{2}-2M}{\ell^{2}}=0,$ $\tilde{Q}$
$=\frac{\sqrt{2}Q}{\ell}$ and $\tilde{q}$ $=\sqrt{\frac{8}{3}}\frac{q}{\ell}$
to get
\begin{equation}
\frac{d^{2}u}{d\psi^{2}}+\tilde{Q}^{2}u\ln u-\tilde{q}^{2}u^{3}=0,
\end{equation}
such that the inverse of the solution of this equation is plotted in the Fig.
(4) (i.e., $r=\frac{1}{u}$ versus $\psi).$ Obviously, from the figure, one
observes that, the roles of Maxwell and YM charges outside the black hole are
in contrast with each other. Next, we consider Eq. (55) and set the mass of
the black hole to be zero. By adjusting charges in terms of the angular
momentum, we express the equation in the form%
\begin{equation}
\frac{d^{2}u}{d\psi^{2}}+\left(  1+\frac{\tilde{Q}^{2}}{2}\right)  u+\tilde
{Q}^{2}u\ln u-\tilde{q}^{2}u^{3}=0.
\end{equation}
One notices that, if $\tilde{Q}=0,$ this reduces to the Duffing type equation
as,%
\begin{equation}
\frac{d^{2}u}{d\psi^{2}}+u-\tilde{q}^{2}u^{3}=0
\end{equation}
which can be considered for small $\tilde{q}$, as a perturbed simple harmonic
oscillator. In our work, however, we are interested to consider both terms
(i.e., Maxwell and YM terms). From Eq. (55) we observe that u=1 forms a
circular orbit provided the condition (with $M\neq0$)
\begin{equation}
Q^{2}-\frac{8}{3}q^{2}=2M-\ell^{2}%
\end{equation}
holds. We investigate the stability of this orbit by choosing
\begin{equation}
u=1+a\cos\beta\psi
\end{equation}
in which $\beta$ and $a$ are constants such that $a\ll1.$ By substitution we
obtain
\begin{equation}
\beta=\pm\frac{\sqrt{2}}{\ell}\sqrt{Q^{2}-\frac{8}{3}q^{2}}%
\end{equation}
so that stability of the circular orbits are attained provided $Q^{2}>\frac
{8}{3}q^{2}.$ Thus, a dominating YM charge gives rise to a deeper well and
stable orbits. The foregoing argument can easily be extended to cover
elliptical orbits as well, which will not be repeated here. Let us note that
the possibilities involved in a complete analysis of the Eq. (55) may reveal
different behaviors as well.

\section{EMYM Black Holes in GB Gravity}

In this section we use our previous ansaetze and find solutions with the GB
term. The new action is modified now as
\begin{equation}
I_{G}=\frac{1}{2}\int_{\mathcal{M}}dx^{N}\sqrt{-g}\left(  R+\alpha
\mathcal{L}_{GB}-\frac{\left(  N-1\right)  \left(  N-2\right)  }{3}%
\Lambda-F_{\mu\nu}F^{\mu\nu}-\mathbf{Tr}\left(  F_{\mu\nu}^{(a)}F^{(a)\mu\nu
}\right)  \right)  ,
\end{equation}
where the new term $\mathcal{L}_{GB}=R_{\mu\nu\alpha\beta}R^{\mu\nu\alpha
\beta}-4R_{\alpha\beta}R^{\alpha\beta}+R^{2}$ is the GB Lagrangian with the
constant GB parameter $\alpha.$ The Maxwell and YM ansatz are chosen as in the
previous section. The EMYMGB equation that helps us to determine $f\left(
r\right)  $ is given by
\begin{gather}
\left(  \Delta-\frac{\Lambda}{3}\left(  N-1\right)  r^{4}\right)  \left(
N-2\right)  r^{\left(  2N-6\right)  }-2\left(  N-3\right)  ^{2}q^{2}r^{4}=0,\\
\Delta=\left(  -r^{3}+2\overset{\sim}{\alpha}r\left(  f\left(  r\right)
-1\right)  \right)  f^{\prime}\left(  r\right)  +\overset{\sim}{\alpha}\left(
N-5\right)  \left(  f\left(  r\right)  -1\right)  ^{2}-\nonumber\\
r^{2}\left(  N-3\right)  \left(  f\left(  r\right)  -1\right)  -Q^{2}\left(
N-3\right)  .\nonumber
\end{gather}
Solution for $f\left(  r\right)  $ follows as%
\begin{equation}
f_{\pm}\left(  r\right)  =\left\{
\begin{array}
[c]{ccc}%
1+\frac{r^{2}}{4\alpha}\pm\Psi & , & N=5\\
1+\frac{r^{2}}{2\overset{\sim}{\alpha}}\left(  1\pm\Upsilon\right)  & , &
N\geq6
\end{array}
\right.
\end{equation}
where%
\begin{gather}
\Psi=\sqrt{1+\frac{M}{2\alpha}+\left(  \frac{\Lambda}{3}+\frac{1}{8\alpha
}\right)  \frac{r^{4}}{2\alpha}+\frac{Q^{2}\ln\left(  r\right)  }{\alpha
}-\frac{2q^{2}}{3\alpha r^{2}}},\\
\Upsilon=\sqrt{1+\frac{4\overset{\sim}{\alpha}\Lambda}{3}+\frac{4\overset
{\sim}{\alpha}Q^{2}\left(  N-3\right)  }{\left(  N-5\right)  r^{4}}%
+\frac{8M\overset{\sim}{\alpha}}{r^{N-1}}-\frac{8\left(  N-3\right)
q^{2}\overset{\sim}{\alpha}}{\left(  N-2\right)  r^{2\left(  N-2\right)  }}},
\end{gather}
in which $\overset{\sim}{\alpha}=\left(  N-3\right)  \left(  N-4\right)
\alpha$ and $M$ is a constant of integration to represent mass. In the limit
$\alpha\rightarrow0$ the expression for $f_{-}\left(  r\right)  $ reduces to
the ones in the previous section , as it should and for positive branch
$\alpha$ can not be zero. For $Q=0$ our result reduces to the one known before
for the EMGB theory \cite{9}. Similarly for $q=0$ we recover the results
obtained previously \cite{10,11,12}. In Fig. (5) we plot the radius of the
event horizon of the 5-dimensional EMYMGB solution in terms of Maxwell and YM
charges. The different effects of these charges we use to emphasize. For
$N\geq6$ it can easily be seen that the Maxwell and YM terms have opposite
signs. It is remarkable to observe that the solutions (64), behave
asymptotically dS (AdS) such that the effective cosmological constant may be
written as%
\begin{equation}
\Lambda_{eff}\tilde{=}\left\{
\begin{tabular}
[c]{lll}%
$-\frac{3}{4\alpha}\left(  1\pm\sqrt{1+\frac{8\Lambda}{3}\alpha}\right)  $ &
, & $N=5$\\
$-\frac{3}{2\overset{\sim}{\alpha}}\left(  1\pm\sqrt{1+\frac{4\Lambda}%
{3}\overset{\sim}{\alpha}}\right)  $ & , & $N\geq6$%
\end{tabular}
\ \ \ \ \ \ \ \ \right.
\end{equation}
in which in the limit of $\overset{\sim}{\alpha}\rightarrow0$, the negative
branch, admits $\Lambda_{eff}\rightarrow\Lambda$ and the positive branch in
the limit of $\Lambda=0$, gives $\Lambda_{eff}=-\frac{3}{\overset{\sim}%
{\alpha}}.$

For the case $N=5$, on the other hand, the range of $r$ determines the sign of
YM term. Although the $\pm$ signs determine the role of both terms we prefer
the choice $\left(  -\right)  $ under which in the limit $\alpha\rightarrow0$
we recover the EMYM black holes. In conclusion, by studying the Maxwell and YM
fields together we see that these fields compete for dominance in dimensions
$N\geq5$ . We observe on the other hand that in lower dimensions $\left(
N=3,4\right)  $ their roles remain indistinguishable. The asymptotic solutions
reveal that the physical results are independent from one charge or the other.
It is our belief\ \ that this may be helpful in understanding the problem of
confinement (i.e., accretion, collapse) versus the electric and YM charges.

\section{EMYMGB solution for a specific Ansatz for $N\geq4$ (BR type)}

In this section we choose our metric ansatz as
\begin{equation}
ds^{2}=-f\left(  r\right)  dt^{2}+\frac{dr^{2}}{f\left(  r\right)  }%
+h^{2}d\Omega_{N-2}^{2},\text{ \ \ }N\geq4,
\end{equation}
in which $h=$ constant, is to be expressed in terms of the parameters of the
theory. While the YM field will be chosen as before, our Maxwell field will be
different. For the present purpose let our Maxwell 1-form be given by the
choice%
\begin{equation}
\mathbf{A}=qrdt,
\end{equation}
where the constant $q$ is related to the electric charge. This choice has the
feature that the only non-vanishing electromagnetic field 2-form%
\begin{equation}
\mathbf{F}=-qdt\wedge dr,
\end{equation}
yields a uniform electric field. The non-vanishing Maxwell energy-momentum
tensor components $T_{\text{Max }b}^{a}$ are
\begin{equation}
T_{\text{Max }b}^{a}=-q^{2}\text{diag}\left[  1,1,-1,-1,..,-1\right]  .
\end{equation}
The non-zero YM energy-momentum components $T_{\text{YM }b}^{a}$ are
\begin{align}
T_{\text{YM }b}^{a}  &  =-\frac{\left(  N-3\right)  \left(  N-2\right)  Q^{2}%
}{2h^{4}}\text{diag}\left[  1,1,\kappa,\kappa,..,\kappa\right]  ,\text{
\ \ \ }N\geq4,\\
\kappa &  =\frac{N-6}{N-2}.\text{ \ \ \ \ }\nonumber
\end{align}
The field equations (4) with the GB and $\Lambda$ terms, on the premise that
$h=$ constant, reduces to
\begin{gather}
\left[  h^{2}+2\alpha\left(  N-3\right)  \left(  N-4\right)  \right]
f^{\prime\prime}-\left(  N-3\right)  \left(  N-4\right)  \left[
1+\alpha\left(  N-5\right)  \left(  N-6\right)  \right]  +\nonumber\\
\frac{1}{3}\left(  N-1\right)  \left(  N-2\right)  \Lambda h^{2}+\frac{\left(
N-3\right)  \left(  N-6\right)  }{h^{2}}Q^{2}-2q^{2}h^{2}=0,\text{
\ \ \ }N\geq4.
\end{gather}
The solution for $f\left(  r\right)  $ can be expressed as
\begin{equation}
f\left(  r\right)  =C_{1}r^{2}+C_{2}r+C_{3}%
\end{equation}
where $C_{2}$ and $C_{3}$ are integration constants while $C_{1}$ is a
constant depending on the parameters of the theory. Explicitly we have%
\begin{align}
C_{1}  &  =\left\{  \frac{1}{2}\left(  N-3\right)  \left(  N-4\right)  \left[
1+\alpha\left(  N-5\right)  \left(  N-6\right)  \right]  -\frac{1}{6}\left(
N-2\right)  \left(  N-1\right)  \Lambda h^{2}\right.  -\nonumber\\
&  \left.  \frac{\left(  N-3\right)  \left(  N-6\right)  Q^{2}}{2h^{2}}%
+q^{2}h^{2}\right\}  /\left[  h^{2}+2\alpha\left(  N-3\right)  \left(
N-4\right)  \right]  ,\text{ \ \ \ }N\geq4.
\end{align}
The constant $h^{2}$ is also expressible by
\begin{equation}
\frac{2}{3}h^{2}=\frac{\left(  N-3\right)  \left(  N-2\right)  \pm
\sqrt{\left(  N-3\right)  \left(  N-2\right)  \left[  K\left(  N-2\right)
+8q^{2}L\right]  }}{6q^{2}+\left(  N-1\right)  \left(  N-2\right)  \Lambda},
\end{equation}
in which we have abbreviated
\begin{align}
K  &  =\frac{4}{3}\left(  N-1\right)  \left[  \left(  N-5\right)  \left(
N-4\right)  \alpha-Q^{2}\right]  \Lambda+\left(  N-3\right)  ,\\
L  &  =\left(  N-5\right)  \left(  N-4\right)  \alpha-Q^{2},\text{ \ \ }%
N\geq4,\nonumber
\end{align}
and $(+)$ and $(-)$ signs are chosen in the Maxwell and YM limits,
respectively. From these expressions we obtain the EMYM\ limit by setting
$\alpha=0$. Similarly the EMGB and EYMGB limits can be obtained by setting
$Q=0$ and $q=0,$respectively. For $C_{2}\neq0\neq$ $C_{3}$ we can have the
roots of $f\left(  r\right)  =0,$ which signals the horizons for black holes.
\ The abundance of parameters in the EMYMGB theory creates a large class of
possibilities admitting various black hole solutions which we shall not pursue
here. By choosing $C_{2}=C_{3}=0$ and constraining $h^{2}=\frac{\beta}{C_{1}%
},$ for a suitable constant $\beta\left(  >0\right)  ,$ followed by a
redefinition of time we cast the line element into
\begin{equation}
ds^{2}=\frac{h^{2}}{\beta}\left(  \frac{-dt^{2}+dr^{2}}{r^{2}}+\beta
d\Omega_{N-2}^{2}\right)  ,
\end{equation}
which is of the static BR form. For $\alpha=0=\Lambda=q$ (as a limit ) we
arrive at \cite{13,14}%
\begin{equation}
ds^{2}=\frac{Q^{2}}{N-3}\left(  \frac{-dt^{2}+dr^{2}}{r^{2}}+\left(
N-3\right)  d\Omega_{N-2}^{2}\right)  .
\end{equation}

In general since $\beta\neq1$ conformal flatness is not satisfied. Only for
$N=4$ we have the case of exact BR which is conformally flat. However in
general we have shown that in the EMYMGB theory we construct a metric which is
of BR type, albeit it fails to satisfy the conformal flatness. In the pure
Maxwell limit, ($q\neq0),$ $Q=\Lambda=\alpha=0$ and adopting $C_{2}=C_{3}=0,$
we obtain (after scaling)
\begin{equation}
ds^{2}=\frac{1}{C_{1}}\left[  \frac{-dt^{2}+dr^{2}}{r^{2}}+\left(  N-3\right)
^{2}d\Omega_{N-2}^{2}\right]  ,
\end{equation}
which is also of similar type with the constant $C_{1}=2q^{2}\left(
\frac{N-3}{N-2}\right)  .$ This is in agreement with the higher dimensional BR
metric in the EM theory \cite{13}. We recall that following the method of
Ginsparg and Perry \cite{15} the N-dimensional YM-BR solution can be expressed
in the form
\begin{equation}
ds^{2}=\frac{Q^{2}}{\left(  N-3\right)  }\left[  -\sinh^{2}\chi dT^{2}%
+d\chi^{2}+\left(  N-3\right)  d\Omega_{N-2}^{2}\right]  .
\end{equation}
The transformation that takes us to this result in the limit $\epsilon
\rightarrow0$ is given by taking%
\begin{equation}
f\left(  r\right)  =\frac{N-3}{Q^{2}}\left(  r+t\right)  \left(  r-t\right)
,\text{ \ \ }t=\frac{Q^{2}}{\left(  N-3\right)  \epsilon}T,\text{
\ \ }r=\epsilon\cosh\chi,
\end{equation}
while the magnetic type YM field remains unchanged.

\section{Energy and Causality Conditions}

\subsection{$N\geq5-$dimensions (RN type)}

The energy conditions (EC) of the matters associated with the energy momentum
tensor given by (12) and (13) for $N\geq5-$dimensions, i.e.
\begin{equation}
T_{b}^{a}=T_{Max\text{ }b}^{a}+T_{YM\text{ }b}^{a}%
\end{equation}
can be studied by using the definition of the energy-density of the matter
$\rho$\cite{18}%
\begin{equation}
\rho=-T_{t}^{t}=-T_{r}^{r}=\frac{\left(  N-3\right)  ^{2}q^{2}}{r^{2\left(
N-2\right)  }}+\frac{\left(  N-3\right)  \left(  N-2\right)  Q^{2}}{2r^{4}},
\end{equation}
the principal pressures $p_{i}$%
\begin{equation}
p_{i}=T_{i}^{i}\text{ \ \ (no sum convention)}%
\end{equation}
and the effective pressure%
\begin{equation}
p_{eff}=\frac{1}{N-1}\sum\limits_{i=1}^{N-1}p_{i}.
\end{equation}

\subsubsection{Weak Energy Condition (WEC)}

The WEC may be expressed as%
\begin{equation}
\rho\geq0\text{ \ \ and \ \ }\rho+p_{i}\geq0,\text{ \ \ (}i=1,2...N-1\text{)}%
\end{equation}
which holds true in any dimensions $N\geq4,$ by the energy momentum tensor
given by (83).

\subsubsection{Strong Energy Condition (SEC)}

The SEC states that%
\begin{equation}
\rho+\sum\limits_{i=1}^{N-1}p_{i}\geq0\text{ \ \ and \ \ }\rho+p_{i}%
\geq0,\text{ \ \ \ (}i=1,2,...,N-1\text{)}%
\end{equation}
which for $4<N\leq6$ holds true, but for $N\geq7$ the SEC is satisfied for
$r\leq r_{\sec}$ in which%
\begin{equation}
r_{\sec}=\left(  \frac{2\left(  N-3\right)  q^{2}}{\left(  N-6\right)  Q^{2}%
}\right)  ^{\frac{1}{2\left(  N-4\right)  }},\text{ \ \ }N\geq7.
\end{equation}

\subsubsection{Dominant Energy Condition (DEC)}

In accordance with the DEC, which are given by%
\begin{equation}
\rho\geq\left\vert p_{i}\right\vert ,\text{ \ \ (}i=1,2,...,N-1\text{)}%
\end{equation}
our energy momentum tensor satisfies these for any dimensions $N>4.$

\subsubsection{Causality Condition (CC)}

We express the CC after knowing that $\rho>0$ as%
\begin{equation}
0\leq p_{eff}<\rho,
\end{equation}
which is satisfied by the energy momentum tensor given by (83) for $N=4,5.$
For $N\geq6$ (91) is satisfied if $r<r_{cc}$ where%
\begin{equation}
r_{cc}=\left(  \frac{2\left(  N-3\right)  ^{2}q^{2}}{\left(  N-2\right)
\left(  N-5\right)  Q^{2}}\right)  ^{\frac{1}{2\left(  N-4\right)  }},\text{
\ \ }N\geq6.
\end{equation}

\subsection{$N=3,4-$Dimensions (RN type)}

In $4-$dimensions, the energy momentum tensor simply reads%
\begin{equation}
T_{b}^{a}=-\frac{q^{2}+Q^{2}}{r^{4}}diag\left[  1,1,-1,-1\right]
\end{equation}
in which WEC, SEC, DEC and CC are all verified. In $3-$dimensions also the
energy momentum tensor which is given by%
\begin{equation}
T_{b}^{a}=-\frac{q^{2}+Q^{2}}{r^{2}}diag\left[  1,1,-1\right]
\end{equation}
satisfies all the energy and causality conditions.

\subsection{$N\geq4-$dimensions (BR type)}

To investigate the energy conditions of second type solutions (BR type) we
rewrite the energy momentum tensor of the system in the form of%
\begin{equation}
T_{b}^{a}=T_{Max\text{ }b}^{a}+T_{YM\text{ }b}^{a}%
\end{equation}
where%
\begin{align}
T_{\text{Max }b}^{a}  &  =-q^{2}\text{diag}\left[  1,1,-1,-1,..,-1\right]  ,\\
T_{\text{YM }b}^{a}  &  =-\tilde{Q}^{2}\text{diag}\left[  1,1,\kappa
,\kappa,..,\kappa\right]  ,\text{ \ \ \ }N\geq4,\text{ }\nonumber\\
\text{\ \ }\kappa &  =\frac{N-6}{N-2}\text{, \ \ }\tilde{Q}^{2}=\frac{\left(
N-3\right)  \left(  N-2\right)  Q^{2}}{2h^{4}}.\nonumber
\end{align}
One can show that the WEC is satisfied for arbitrary $N\geq4.$ The SEC is also
verified for $N=4,5,6$ but for $N\geq7,$ only under the condition
\begin{equation}
q^{2}\geq\kappa\tilde{Q}^{2}%
\end{equation}
the SEC is satisfied. It is also easy to show that, DEC is satisfied for any
dimensions. Finally, the CC is satisfied for $N=4,5,$ but for $N\geq6$ it
becomes valid with the additional condition%
\begin{equation}
q^{2}\geq\frac{N-5}{N-3}\tilde{Q}^{2}.
\end{equation}

\section{Conclusion}

Our exact solution in the first part of the paper suggests that for $N\geq5$,
the Maxwell and YM fields compete for dominance in the asymptotic regions.
That is, for $r\rightarrow0$ ($r\rightarrow\infty)$ the Maxwell charge q (the
YM charge Q) dominates. This may shed light on the problem of gravitational
confinement (i.e., accretion, collapse) versus the Maxwell and YM charges. As
a drawback of our model the YM field is treated, in analogy with the Maxwell
field, entirely classical. The Newtonian approximation in the polar plane
consisting of the coordinates $\left(  r,\psi\right)  $ reveals that the YM
charge deepen the potential well to form bound states. In lower dimensions
(i.e., $N=3,N=4$), however the roles of q and Q remain indistinguishable. The
presence of a logarithmic term for $N=3$ and $N=5$ is a distinctive property
compared to other dimensions. An effective cosmological constant can be
defined from the GB parameters for $r\rightarrow\infty.$In the last part of
the paper where we introduced a different ansatz, we present exact solution to
the EMYMGB theory which can represent a variety of black holes. Another
possibility is by the choice of parameters to cast the metric into the static
BR form which lacks conformal flatness but may be important in
supergravity/string theory\cite{16}, as well as in cosmology. Finally,
validity of the energy/ causality conditions are discussed for all solutions
that are obtained in the paper.

\section{Figure captions:}

Fig. (1): Radius of the event horizon of the 5-dimensional, EMYM black hole in
terms of the Maxwell and YM charges, and specific values for $M$ and $\Lambda
$. The non-black hole region is shown as NBH.

Fig. (2): A plot of $T_{H}$ for 3-dimensional EMYM black hole solution, in
terms of $\Lambda$ and $\left(  Q^{2}+q^{2}\right)  .$ We set the mass of the
black hole to be unit.

Fig. (3): The effective potential versus $r$. We aim to compare the roles of
Maxwell and YM charges, versus their effects in the $V_{eff}\left(  r\right)
$ in 5-dimensions$.$ It is seen that it is the YM charge Q which provides deep
potential well apt for the bound states.

Fig. (4): A plot of radius of a particle whose initial values are set as:
$r(\psi=0)=1,\left.  \frac{dr}{d\psi}\right\vert _{\psi=0}=-0.7$. For
$\tilde{Q}=3$ and $\tilde{q}<1.215953$ the orbit of the particle is closed and
it oscillates around $r=1,$ but for $\tilde{q}\geq1.215953$ the orbits are not
closed and the particle falls toward r=0. We note that the mass m is chosen in
accordance with $1+\frac{Q^{2}-2m}{\ell^{2}}=0.$

Fig. (5): Radius of the event horizon of the 5-dimensional, EMYMGB black hole
in terms of Maxwell and YM charges, and specific values for $M$ , $\alpha$ and
$\Lambda$. The non-black hole region is shown as NBH. (These plot may be
compared with those in Fig. (1)).


\begin{thebibliography}{99}                                                                                               %


\bibitem {1}N. Okuyama and K. I. Maeda, Phys. Rev. D \textbf{67} (2003) 104012.

G.W. Gibbons and P.K. Townsend, Class. Quant. Grav. \textbf{23} (2006) 4873.

\bibitem {2}Y. Brihaye, E. Radu and D. H. Tchrakian, Phys. Rev. D \textbf{75}
(2007) 024022.

\bibitem {3}S. H. Mazharimousavi and M. Halilsoy, Phys. Lett. B \textbf{659}
(2008) 471.

\bibitem {4}C. W. Misner, K. S. Thorne and J. A. Wheeler, \emph{Gravitation}
(Freemann, San Fransisco, 1973).

\bibitem {5}T. T. Wu and C. N. Yang, in \emph{Properties of Matter Under
Unusual conditions},edited by H. Mark and S. Fernbach (Interscience, New
York,1969), p. 349.

\bibitem {6}P. B. Yasskin, Phys. Rev. D \textbf{12} (1975) 2212.

\bibitem {7}M. Ba\~{n}ados, C. Teitelboim and J. Zanelli, Phys. Rev. Lett.
\textbf{69} (1992) 1849.

C. Mart\'{\i}nez, C. Teitelboim and J. Zanelli, Phys. Rev. D \textbf{61}
(2000) 104013

D. Ida, Phys. Rev. Lett. \textbf{85} (2000) 3758.

\bibitem {8}R. M. Corless, G. H. Gonnet, D. E. G. Hare, D. J. Jeffrey and D.E.
Knuth. Adv. Comput. Math 5 (1996) 329.

\bibitem {9}M. H. Dehghani, Phys. Rev. D \textbf{70} (2004) 064019.

\bibitem {10}S. H. Mazharimousavi and M. Halilsoy, Phys. Rev. D \textbf{76}
(2007) 087501.

S. H. Mazharimousavi and M. Halilsoy, Phys. Lett. B \textbf{665} (2008) 125.

\bibitem {11}Yves Brihaye, A. Chakrabarti, Betti Hartmann and D. H. Tchrakian,
Phys. Lett. B \textbf{561} (2003) 161.

\bibitem {12}A. Chakrabarti and D. H. Tchrakian, Phys. Rev. D\textbf{\ 65}
(2001) 024029.

\bibitem {13}V. Cardoso, O. J. C. Dias and J. P. S. Lemos, Phys. Rev. D
\textbf{70} (2004) 024002.

\bibitem {14}S. H. Mazharimousavi and M. Halilsoy, "\emph{Dilatonic black
holes and Bertotti-Robinson space times for Yang-Mills Fields}". (2008)
arXiv:0802.3990 [gr-qc].

\bibitem {15}P. Ginsparg and M. J. Perry, Nucl. Phys. B \textbf{222} (1983) 245.

\bibitem {16}D. G. Boulware and S. Deser, Phys. Rev. Lett. \textbf{55} (1985) 2656.

\bibitem {17}M. Nozawa and H. Maeda, Class. Quant. Grav. \textbf{25} (2008)
055009 .

T. Torii, and H. Maeda, Phys.Rev. D \textbf{71} (2005) 124002.

T. Torii, and H. Maeda, Phys.Rev. D \textbf{72} (2005) 064007.

\bibitem {18}M. Salgado, Class. Quant. Grav. \textbf{20} (2003) 4551.

R. M. Wald, General Relativity, University of Chicago Press, Chicago (1984).

\bibitem {19}A. Al-Badavi and M. Halilsoy, Il Nuovo Cimento, \textbf{119 B}
(2004) 931.
\end{thebibliography}
\end{document}